\documentstyle[11pt,moriond,epsfig]{article}

\def\be{\begin{equation}}
\def\ee{\end{equation}}
\def\bea{\begin{eqnarray}}
\def\eea{\end{eqnarray}}

\begin{document}

\vspace*{4cm}

\vspace*{4cm}

\title{LEPTON MIXING: SMALL, LARGE, MAXIMAL?}

\author{A. Yu. Smirnov\,
\footnote{Invited talk given at XXXIV Rencontres de Moriond, 
Electroweak Interactions and Unified Theories, March 13 - 20,  
1999, Les Arcs, Savoie, France.} 
}

\address{International Center for Theoretical Physics,\\ 
Strada Costiera 11, 34100 Trieste, Italy, and  \\ 
Institute for Nuclear Research, RAS, Moscow, Russia}

\maketitle\abstracts{The SuperKamiokande data on atmospheric neutrinos 
imply that $\nu_{\mu}$  has large (or even maximal) 
mixing. It is still open question whether this mixing 
is the flavor one  or mixing with singlet state 
(sterile neutrino). Several tests exist to establish the channel 
of atmospheric neutrino oscillations.  
Large mixing can be a property of the 
second and third generations, so that   
the scheme with single large mixing is realized. 
It can be  a generic property of all 
leptons, thus leading to   the  bi-large  or threefold large mixing
schemes. 
The ambiguity   will be resolved by  identification 
of solution of the solar neutrino problem. 
In this review we  consider phenomenology of various mixing 
schemes.  The key measurements  
 which will allow us to make significant step in  reconstruction of  
the neutrino mass and flavor spectrum are discussed. 
}  

\newpage

\section{Introduction}     

During 736 days of observations the SuperKamiokande (SK)  collaboration 
has detected more than  6300 atmospheric neutrinos \cite{SK}. Other
detectors have recorded  about  2000 
events~\cite{kam,imb,soudan,macro,baksan}.  
With this results we are entering new stage of  
high statistics and high precision studies. 
Events  detected so far differ by  
 topology ($\alpha$): one (Cherenkov) ring $e-$like and 
$\mu-$ like events, two ring events (where the so called $\pi^0$ 
events are of special interest), multi-ring events, 
upward-going muons etc.. The energy, $E$, and the zenith angle, 
$\Theta$, dependences of events of each topology 
have been measured. That is, numbers of events, 
$N_{\alpha} (E, \Theta)$, 
of the type $\alpha$ have been found  in various 
energy and  zenith angle bins.

With increase of statistics  various data from the SuperKamiokande as
well as from 
SOUDAN \cite{soudan} and   MACRO \cite{macro} show better 
consistency. Their implications converge to the same 
oscillation interpretation. 
In particular, the SOUDAN observes the up - down asymmetry of the 
tracks (muons). The best fit of the MACRO data on through-going 
muons corresponds to oscillation interpretation;    
comparison of stopping to through-going muons further confirms 
the oscillations. 
There is also better internal consistency of the SK data themselves. 
One change is rather significant: 
the rate of the $e$-like events has decreased in the last series of
observations,   and correspondingly, 
the double ratio:
${(R_{\mu/e})_{exp}}/{(R_{\mu/e})_{MC}}$
of  the experimental and the simulated 
$\nu_\mu$ to $\nu_e$ ratios $(R_{\mu / e})$ 
has increased. 
Further independent confirmation of the oscillations has been obtained 
from studies of the sub-dominant samples of events 
(two rings events, stopping/through-going etc.). 
The observation of the East - West (geomagnetic) effect~\cite{eastwest} 
confirms overall picture of generation of the 
atmospheric neutrinos.\\ 

On the basis of this 
information we can say with high confidence level that 

\begin{itemize}

\item
The atmospheric {\it muon neutrinos oscillate}: 
$\nu_{\mu} \leftrightarrow \nu_x$, where 
$\nu_x$ is in general a combination of the muon, 
electron, and probably,  sterile neutrinos.

\item
The {\it oscillations are induced by neutrino mass} 
difference and vacuum mixing. Other interpretations~\cite{decay,fcnc} are 
disfavored~\cite{lipari,lisif}. 

\item
The electron neutrino $\nu_e$ 
is not  a dominant component of $\nu_x$. Data favor 
$\nu_{\mu} \leftrightarrow \nu_{\tau}$ as the main (dominating) 
mode of oscillation, although $\nu_{\mu} \leftrightarrow \nu_{s}$,  
as the dominant channel,   is not excluded. 

\end{itemize}

With high statistics  we are now in position 
to make next step: to study sub-leading effects. 
We can approach  the  following three problems:

1. Identification of the channel of oscillations.

2. Clarification of the  role of the $\nu_e$ in oscillations 
of the atmospheric neutrinos.   
In particular, 
determination of the admixture of the electron neutrino 
in the heavy mass eigenstate, $U_{e3}$,    
can be done by 
studies of the multi-GeV $e$-like events.  

3. Searches of oscillation effects driven by smallest $\Delta m^2$
splitting responsible, presumably, for the solution of the solar neutrino
problem. 
This can be done   by precision studies of 
the sub-GeV $e$-like events.

The above effects  are typically smaller than 10\%  
of the main oscillation effect and  samples of events sensitive 
to these effects are an order of magnitude smaller than total number of 
events.\\

Let us outline  situation with solar neutrinos. 
After 708 days of observations the SK has detected 
more than 11500 events~\cite{SKsol}. This opens a possibility to search 
for  {\it signatures} of various oscillation solutions  
in the ranges of oscillation parameters determined  by the  fit of
total rates. That is,  the total event rates in  all existing experiments. 
(This means that we have indeed started crucial checks of proposed 
solutions.) Among these signatures are: \\ 

(i) Distortion of the recoil electron spectrum 
$R(E) \equiv N(E)^{obs}/N(E)^{SSM}$. 
Some  distortion (deviation of $R(E)$ from a constant) 
has already been observed. The main feature 
of the data is that below electron energy 13.5 MeV they are in agreement
with undistorted spectrum. Above 13.5 MeV there is about  3 $\sigma$ 
excess whose interpretation is still unclear:  It can be
just statistical fluctuation, or it can be due to large 
flux of the $hep$-neutrinos, or due to  effect of the vacuum 
oscillations with relatively large $\Delta m^2$.  
It will be  important  to see the result on the spectrum in the lowest 
bin 5.0 - 5.5 MeV. 

(ii) The night-day asymmetry 
\begin{equation} 
A_{N/D} \equiv 2\frac{N - D}{N + D} \approx \frac{N}{D} - 1, 
\label{dn}
\end{equation}
where $N$ ($D$)  is the nighttime (daytime) signal integrated over 
energies above 6.5 MeV  and averaged over the year. 
The asymmetry is observed at the level 6\% which differs 
from zero by 1.6 $\sigma$. Recent data from 825 days of
observations~\cite{Kajita} 
has  increased this significance to 2$\sigma$. 

(iii) The zenith angle distribution of events averaged over the year:  
$\bar{F}(\Theta)$. 

(iv) Seasonal variations of the signal. One can characterize it 
by the winter-summer asymmetry: 
\begin{equation}
A_{W/S} \equiv 2\frac{W - S}{W + S}~,  
\label{season}
\end{equation}
where $W$ and $S$ are the signals averaged over winter time 
(November 15 - February 15) and summer time 
(May 15 - August 15) respectively. 
Important criteria can be obtained from the  dependence 
of the asymmetry  on the energy threshold. In fact, the SK sees strong
(but statistically insignificant) enhancement of the 
variations with increase of the threshold. 
Being confirmed, it will allow to  identify  
the  solution.\\ 

The main  conclusion from analysis 
of present situation is that {\it different  datasets  favor different 
solutions of the problem}. In particular,  
the N/D asymmetry and  the zenith angle dependence 
indicate  Large Mixing Angle (LMA) MSW solution. Seasonal variations and
the spectrum distortion 
testify for the Vacuum Oscillation (VO) solution, 
whereas the total rates of signals prefer  Small Mixing Angle 
(SMA) MSW solution. 
The significance of all these indications is about  $2\sigma$,  
and clearly, more data is needed to make definite conclusion. \\

Concerning the third, LSND~\cite{LSND}, evidence of the neutrino mass and
mixing,  it seems that  the oscillation interpretation of the 
LSND result can not be completely excluded by  KARMEN~\cite{KAR}.  
Situation could be changed if 
the KARMEN will start to see some excess of events. 
However,  in any case new experiments, like  the MiniBoon~\cite{miniboon}, 
are needed,

Other key results in the ``game"  are:  
(i) the bounds on neutrino mass and mixing which follow from  reactor 
experiments CHOOZ~\cite{CHOOZ} and Palo-Verde~\cite{paloverde},  
(ii) new bound on the effective Majorana mass from the 
neutrinoless double beta decay~\cite{baudis}  
and (iii)  bound on  neutrino masses inferred by  the large 
scale structure of the Universe. \\

There are two related  fundamental goals of  these studies: 
\begin{itemize}

\item Reconstruction of the neutrino mass spectrum and lepton mixing. 
Clearly,  it is as fundamental as reconstruction of the 
quark mass spectrum and determination of the quark mixings. 

Eventually, it will be of great importance to compare quark and
lepton spectra and mixings. This will give certain insight to  

- unification of the quarks and leptons; 

- quark - lepton symmetry; 

- mechanism of neutrino mass generation; 

- origin of the fermion masses; 

- new symmetries  and  new mass (energy) 
scales in Nature.

\item 
Searches for new neutrino states, singlets 
of the standard model group. 
Although these new light fermions can couple in some way 
({\it e.g.} via non-renormalizable interactions) with other  SM
particles,  their presence can be revealed in 
the neutrino physics only  
due to unique properties of neutrinos. 

\end{itemize}

There are two main  motivations for introduction of the 
sterile neutrinos: the first one is  to explain all neutrino anomalies 
including the LSND result in terms 
of neutrino oscillations. The second is  to explain the large
lepton mixing itself (see below). 

Notice that even if sterile neutrinos do not play major role in 
explanation of the neutrino problems, they may produce 
some sub-leading effects. And  searches for 
possible presence of  
sterile neutrinos is of great importance. Existence 
of such  fermions can lead us far beyond the Standard Model. 
Practically all extensions of the Standard model 
predict existence of the singlets. 
Open questions are their parameters: 
masses and mixing. The presence 
of sterile neutrinos with parameters which can 
give observable effects  
is rather plausible. Indeed, 
 even Planck scale effects can produce the mass terms  
which lead to effects can be seen in solar and supernove
neutrino fluxes~\cite{planck}. 
The mixing mass term of singlet fermion with neutrinos can be of the
order~\cite{BenSmi}  
\begin{equation} 
\bar{m} = \frac{v M}{M_{Pl}} \sim 10^{-4} 
\left(\frac{M}{1 {\rm TeV}}\right) {\rm eV}~,  
\end{equation}
where $v$ is the vev of the higgs doublet,  
the  mass $M$ can be 
the gravitino mass in  of the SUGRA, $m_{3/2}$~\cite{BenSmi}, 
or the $\mu$ parameter of the SUSY~\cite{Karim,DvaliNir}, or  
the scale of breaking 
of an  additional gauge $U(1)$ factor~\cite{BenSmi,Lang} etc.. 
Clearly,  the mass scale in the denominator smaller than   
of $M_{Pl}$ (which would imply  new physics below the Planck scale)  
will produce larger mixing mass. Of course, crucial 
question   is the one  about  mass of the singlet. 
The mass of the order ${ M^2}/{M_{Pl}}$ 
can lead to observable effects.

\section{Small }

\begin{figure}[ht]
\begin{center}
\epsfig{figure=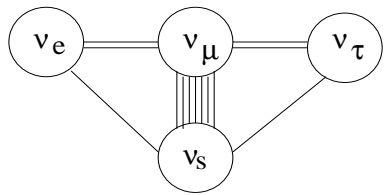,width=5cm,height=2.5cm}
\end{center}  
\caption{The small flavor mixing scheme. The number of lines 
which connect circles reflect a strength of mixing. 
\label{fig:1}}
\end{figure}

\begin{figure}[htb]
\hbox to \hsize{\hfil\epsfxsize=8cm\epsfbox{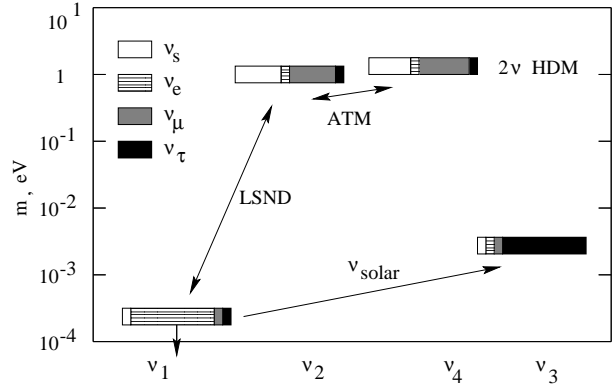}\hfil}
\caption{~~The neutrino masses and mixing
scheme with small flavor mixing and one sterile neutrino.
Boxes correspond to the mass eigenstates. The sizes of different
regions in the boxes show admixtures of different flavors. Weakly hatched
regions correspond to the electron flavor, strongly hatched regions depict
the muon flavor, black regions present the tau flavor.  
}
\label{fgu}
\end{figure}

The only possibility to explain the atmospheric neutrino data and still to
keep small  flavor mixings 
(similar to quark mixings) is to introduce the sterile neutrino 
which mixes strongly with $\nu_{\mu}$, so that the 
atmospheric neutrinos undergo 
$\nu_{\mu} \leftrightarrow \nu_s$ oscillations (fig. \ref{fig:1}). 
In this case the solar neutrino problem should be  solved 
by the small mixing angle MSW conversion. 
There are two versions  
(in the schemes with only one sterile neutrino):   
$$
\nu_{e} \rightarrow \nu_{\tau}
$$     
which corresponds to the scheme~\cite{PTV} shown in  fig. \ref{fgu} , or 
$$
\nu_{e} \rightarrow \nu_{\mu}, \nu_{s}~. 
$$
In the first case one can also explain
the LSND result. The second case is characterized by mass hierarchy 
$m_1 \ll m_2 \ll m_4 \ll m_3$, where the heavy state 
$\nu_3 \approx \nu_{\tau}$ can be in the eV- range thus 
contributing significantly to the HDM.  $\nu_{\mu}$ and 
$\nu_s$ are strongly mixed in $\nu_2$ and $\nu_4$ 
whose masses are determined the solar and the atmospheric mass scales. 
(see~\cite{LS} for details). 

The key question is why sterile neutrino mixes strongly with $\nu_{\mu}$ 
and its mixing with other neutrinos either absent or  very small. 
The suppression of the $\nu_{e} -  \nu_{s}$ and  $\nu_{\tau} -  \nu_{s}$ 
mixing can be related to  hierarchy of the diagonal mass terms of the 
active neutrino components and closeness of the sterile neutrino mass 
to the mass of $\nu_{\mu}$.

Let us consider generic  signatures of the schemes 
which employ the  
$\nu_{\mu} \leftrightarrow \nu_s$ solution of the 
atmospheric neutrino 
problem and small mixing MSW solution of the solar neutrino problem. 
The $\nu_{\mu} \leftrightarrow \nu_s$ solution 
(see {\it e.g.} Ref. ~\cite{atmster}) 
has two main features as 
compared with $\nu_{\mu} \leftrightarrow \nu_{\tau}$~\cite{}. 

1). The neutral current interactions are suppressed 
in the $\nu_{\mu} \leftrightarrow \nu_s$ case  
which can be realized  by studying the reaction~\cite{VS,suzuki}
\begin{equation}
\nu N \rightarrow \nu N \pi^0.
\end{equation}
At the SK, ``$\pi^0$" - events can also be  generated by the charged
current reactions, {\it e.g. }, 
$\nu_{\mu} N \rightarrow \mu  N' \pi^0$,  when muon  is below the   
Cherenkov threshold (similar is for electron neutrinos).
The $\nu_{\mu} -\nu_{\tau}$ oscillations only weakly
suppress the number of ``$\pi^0$",
whereas $\nu_{\mu} -\nu_s$ can suppress  the
``$\pi^0$" rate by  20 - 30 \%. 
Preliminary SK data for 736 days give 279 ``$\pi^0$" events.
This number even exceeds the expected number of  events.
To avoid  normalization uncertainties one can consider the
ratio of numbers of the ``$\pi^0$" - events and the $e$-like events:
$\pi^0/e$. The experiment gives \cite{SK}
\begin{equation}
\frac{(\pi^0/e)_{data}}{(\pi^0/e)_{MC}} = 1.11 \pm 0.06 (stat) 
\pm 0.02 (MC~stat) \pm 0.26
(syst)
\label{dratio}
\end{equation}
which is consistent with both channels of oscillations.
Large systematic error in (\ref{dratio}) is related to uncertainties in
the cross-sections. Notice that the multi-pion production reactions
give significant contribution to the
$\pi^0$ events (due to Cherenkov radiation threshold one or even more
pions are not detected). In \cite{VS} the total uncertainty
was estimated as being at the level 30\%.
The uncertainty will be diminished by direct measurements
of the cross-section in the ``forward" detector of the long baseline
experiment K2K~\cite{suzuki}.
Another uncertainty  is related to  
background, {\it e.g.},  from  interactions of neutrons
in the detector. For the $\pi^0$ events this background
is much more significant than for the  $e$-like events
since only in 17\% of cases $\pi^0$ will induce the $e$-like event.

It is also possible to study the zenith angle
dependence of the $\pi^0$-events which is free of the
uncertainties in the cross-section~\cite{pakw}. 
Another suggestion is to study the up-down asymmetry  of the
inclusive multi-ring events~\cite{HM}.

Of course, a detection of the tau leptons produced
by the $\nu_{\tau}$ would be direct way to
identify the solution. However, the number of the expected events is
rather small \cite{HM1}, and it is difficult  to identify  them.
The detection of $\nu_{\tau}$ is the main issue of 
future  long base-line experiments.\\ 

2). The zenith angle distributions of the high energy events 
are different for $\nu_{\mu} \leftrightarrow \nu_{\tau}$ and  
$\nu_{\mu} \leftrightarrow \nu_s$
due to matter effect in $\nu_{\mu} \leftrightarrow \nu_s$ 
channel~\cite{LS,LMS,LL}. The  $\nu_{\mu} \leftrightarrow \nu_s$ 
gives   flatter $\Theta$-distribution. 
Matter suppresses  oscillations of high energy neutrinos. 
For the neutrinos whose trajectories cross the mantle 
of the Earth one  expects the strongest oscillation effect 
at $|\cos \Theta| =  0.4 - 0.5$ and rather weak oscillation effect for the
upward-going 
muons with $|\cos \Theta| > 0.7$. 
The  enhancement of oscillations  occurs for the core crossing
trajectories  in contrast with 
naive expectation of further suppression of oscillations   
due to larger density of the core. The enhancement is related to the
parametric effect 
based on the fact that the phase of oscillations acquired in the mantle
and in the core is about $\pi$. 
As the result the ratio (DATA/MC) has two shallow minima at 
$|\cos \Theta| = 0.4  - 0.6$ and at $|\cos \Theta| = 0.85 - 0.95$  
with minimal oscillation effect  at $\cos \Theta \approx 0.8$.  
In contrast, the 
$\nu_{\mu} \leftrightarrow \nu_{\tau}$ oscillations 
 lead to monotonous increase of the suppression with $|\cos \Theta|$. 
For the same values of oscillation parameters the 
$\nu_{\mu} \leftrightarrow \nu_{\tau}$ oscillation 
effect at $\cos \Theta = - 1$ can be two times larger than 
in the case of $\nu_{\mu} \leftrightarrow \nu_s$ oscillations, 
if both are normalized to the same value at 
$\cos \Theta = 0$. 
The  SK data prefer $\nu_{\mu} \leftrightarrow \nu_{\tau}$  
mode, and  $\nu_{\mu} \leftrightarrow \nu_{s}$  
mode is disfavored at $2\sigma$ level~\cite{Kajita}.\\

Main signatures of the SMA solution of the solar neutrino problem are:  

1). The distortion of the recoil electron spectrum which can be
characterized by a unique 
parameter: the slope of the ratio $R(E)$: 
$s  \equiv d (\ln R) / d E$. 
No sharp enhancement of the signal is expected  at the 
high energy  part of the
spectrum unless large  flux of the $hep$ -  neutrinos is introduced. 
In the lowest energy  bins one expects slight turn down of the spectrum.

2). Strong regeneration effect~\cite{graph} is expected for neutrinos
whose trajectories
cross the core~\cite{coreenh} due to the parametric 
enhancement of oscillations~\cite{ETC,Akh,plam} as it was realized 
first in~\cite{Pet} see also~\cite{Akh1}. As the result for $\Delta m^2 >
10^{-2}$ eV$^2$ the N/D asymmetry is mainly due 
to  the signal in the vertical bin 
(fifth night bin N5: $\cos\Theta = - 0.8 \div  -1.0$) $F_5$. 
Using this feature  one can find  the correlation between the relative 
excess of the flux in the N5 bin and the night-day asymmetry: 
\begin{equation}
\frac{F_5}{\bar{F}} -1 \approx 5 A_{N/D}~, 
\end{equation}
where $\bar{F}$ is the average flux during  24 hours  
and $F_5$ is the flux in the fifth night bin. 
Thus,  precise measurements of ${F_5}/{\bar{F}}$ and 
$A_{N/D}$ can give important test of the solution.  

No significant enhancement of signal has been yet observed in the 
N5 bin:  This can be reconciled with result on the  
N/D asymmetry at   $2\sigma$ level. Notice that  in the best fit 
point of the total rates  one predicts rather small 
values of the ${F_5}/{\bar{F}} -1 \sim (1 - 5) \%$ and 
$A_{N/D} \approx (0.2 - 1.0) \%$. 

3) The Earth regeneration effect  leads  to 
seasonal variations of the
flux~\cite{seasonMSW,holanda}  on the top of the
seasonal variations due to the eccentricity of the
Earth orbit (geometrical effect). In the first approximation 
the seasonal dependence can be 
understood  taking into account that regeneration occurs mainly when 
neutrinos cross the core. In this case the seasonal effect is
proportional to weight, $\Omega$,  of the trajectories which cross the
core. 
For the SK place one gets $\Omega_S = 0$ in the summer,  and the 
winter-summer asymmetry can be written as 
\begin{equation}
A_{W/S} = \Omega_W \left(\frac{F_5}{\bar{F}} - 1 \right)  
 = 5 \Omega_W   A_{N/D}~.  
\end{equation}
Notice that in contrast with geometrical effect the regeneration is very
small in the wide interval from March to October  and the effect increases 
in the winter time~\cite{holanda}. This differs also 
from predictions of seasonal asymmetry in the case of the LMA solution  
(see below).  
No significant enhancement of variations with the threshold is expected.

\section{Large }    

\begin{figure}[ht]
\begin{center}
\epsfig{figure=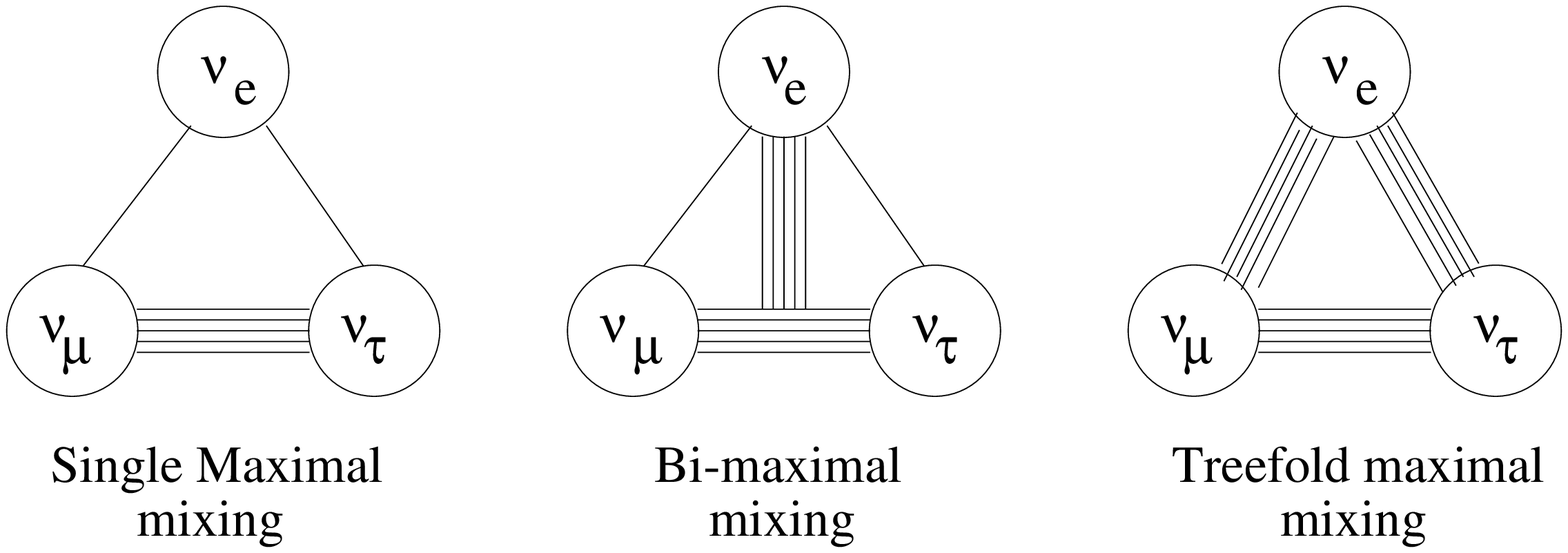,width=11cm,height=3.5cm}
\end{center}  
\caption{The large flavor mixing schemes.
\label{fig:2}}
\end{figure}

There are three different schemes with large mixing 
of three active neutrinos (fig. \ref{fig:2}).  

(i) Scheme with single large (maximal) mixing where only 
$\nu_{\mu}$ and  $\nu_{\tau}$ are mixed strongly and mixings of the 
$\nu_{e}$ flavor are  small. In this case the solar neutrino problem 
can be explained by the SMA MSW conversion. 

(ii) Bi-large (bi-maximal~\cite{bimax}) scheme  when
$\nu_{\mu}$ and $\nu_{\tau}$ are
mixed strongly in the heavy mass eigenstate and 
$\nu_{e}$ is strongly mixed with the orthogonal combination 
of $\nu_{\mu}$ and  $\nu_{\tau}$ in two light eigenstates . 
In this case the solar neutrino problem can be solved by the 
LMA MSW solution or by Vacuum Oscillations.  
The extreme case when both 
mixings are maximal  is called the bi-maximal scheme.

(iii) Threefold  maximal scheme~\cite{three}: All three neutrino flavors
are mixed  equally in the three mass eigenstates.

Let us consider  phenomenological consequences of these schemes. 
Clearly, establishing the solution of the solar neutrino problem will 
play crucial role  in identification of the mixing scheme. 

\subsection{Single Large (Maximal) Scheme} 

\begin{figure}[htb] \hbox to
\hsize{\hfil\epsfxsize=7cm\epsfbox{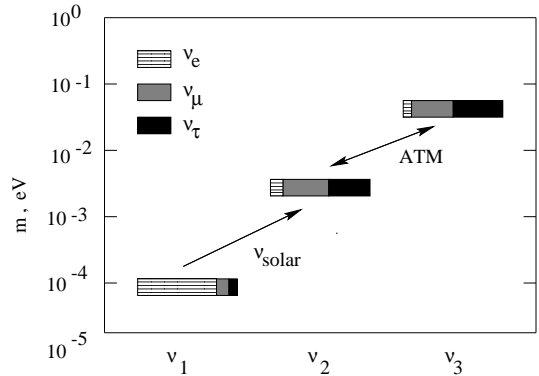}\hfil}
\caption{~~Neutrino mass spectrum with single large mixing. 
}
\label{solat}
\end{figure}

The spectrum is shown in fig. \ref{solat}. 
Generic features  of the scheme  are 
$\nu_{\mu} \leftrightarrow \nu_{\tau}$ oscillations of the 
atmospheric  neutrinos and the 
$\nu_e  \rightarrow \nu_{\mu}/ \nu_{\tau}$ MSW conversion of the 
solar neutrinos. 
Signatures of these solutions have been already discussed 
in  sect. 2. Namely, one predicts  (i) no deficit of 
$\pi^0$-events (ii) sharp $\Theta$-dependence 
of the upward-going muons; (iii) distortion 
of the recoil electron spectrum produced by the boron neutrinos 
from the Sun with positive slope, etc..  

To explain the LSND result one should introduce a 
sterile neutrino. 

\subsection{Bi-large mixing}

The solar neutrino problem can be solved either by the Vacuum oscillations 
or by the LMA MSW conversion. 

\begin{figure}[htb] 
\hbox to \hsize{\hfil\epsfxsize=7cm\epsfbox{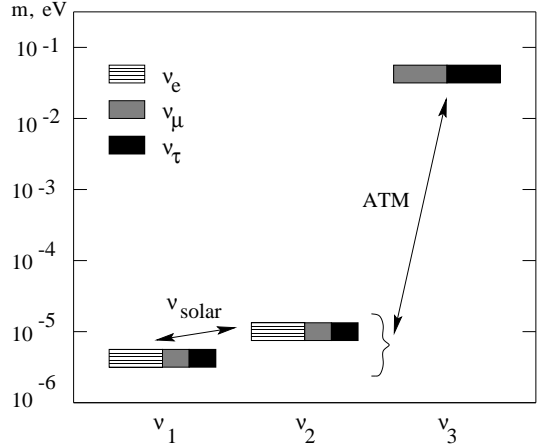}\hfil}
\caption{~~Neutrino masses and mixing pattern  of the  bi-maximal
mixing  scheme.
}
\label{fbimax}
\end{figure}

In the case of the VO solution (the spectrum is shown in fig. 
\ref{fbimax}) the fit of the total event rates  
from  existing experiments determines several disconnected 
regions of the oscillation parameters which can be classified by the phase
of oscillations 
acquired  by boron neutrinos 
with typical energy 10 MeV on the way 
from the Sun to the Earth. The best fit of the  recoil 
electron spectrum can be obtained  in the region with 
$\phi =  5 \pi/2$: $\Delta m^2 \approx 4 \cdot 10^{-10}$ eV$^2$ 
and $\sin^2 2\theta = 0.8 - 1$. 

Typical shape of the recoil electron spectrum distortion $R(E)$ 
is (i) the positive slope  at $E > 11$ MeV which can explain the  excess 
of events  (ii) shallow minimum at $E \sim  8 - 10 $ MeV and 
(iii) weak increase of the $R(E)$ 
with decrease of the energy at $E <  8$ MeV, 
$R(E)$  approaches the constant value  due to strong averaging effect. 
In the last series of observations a significance of the excess has
further decreased which implies   shift of the best fit point 
to  smaller mixing angles: $\sin^2 2\theta = 0.8 - 0.9$.  
This,  however,  worsen  the fit of total rates.  

Reasonable agreement can be also achieved  in the 
``$\pi/2$ -  region"  with $\Delta m^2 \approx 8 \cdot 10^{-11}$ eV$^2$.

(ii) No Night-Day asymmetry and the zenith angle dependence 
of signal should be observed. 
The Earth regeneration effect is negligible.

(iii) The seasonal asymmetry is however expected on the top 
of pure geometrical effect ($1/L^2$) due to eccentricity 
of the Earth orbit and the dependence of
the oscillation probability on 
distance. It was noticed that  seasonal
variations~\cite{gribov,KP} are correlated 
with spectrum distortion~\cite{correl}~\cite{GKK}. If the distortion 
in some part of the spectrum is characterized by the slope 
$s \equiv d\ln R(E)/dE$,  the seasonal asymmetry of the events  in
the same part of the spectrum is proportional to: 
\begin{equation}
A_{W/S} \propto s~. 
\end{equation}
In particular, a positive slope corresponds to positive asymmetry which
is in phase with pure geometrical effect ($1/L^2$). So that 
the  geometrical effect will be enhanced. Negative slope 
corresponds to negative asymmetry and compensation of the  
geometrical  and oscillation effects. The $5\pi/2$-solution predicts
almost flat distortion
below 11 MeV and substantial distortion with a positive slope 
above 11 MeV. Therefore, the seasonal variations should be observed 
in the high energy part of the spectrum. This in turn,  means that 
the relative asymmetry should increase with increase 
of the energy threshold~\cite{SKsol,BFL,seasonalVO}. As we marked, the
data indicate such 
an enhancement, although the conclusion is  statistically insignificant. 

In the case of $``\pi/2$ - region" one can expect 
a turn up of the
spectrum at low energies (negative slope). Consequently,    
the negative asymmetry (due to oscillations) should be observed for the 
low energy events and the positive asymmetry -- for the high 
energy events. 

In ``$5\pi/2$-domain" the oscillation length 
for the beryllium neutrinos ($E \sim 0.86$ MeV) 
is comparable with the change of the distance between the sun and the
Earth during the year $\Delta L$, so that the change of the phase is 
\begin{equation}
\Delta \phi = 2 \pi \frac{\Delta L}{l_{\nu}} \sim 2 \pi. 
\end{equation}
This means that  one should observe  about two periods of 
variations during the year  and  maxima can be in March-April 
and Sept- October~\cite{BFL}. These variations can be observed in  
experiments sensitive to the 
beryllium neutrino flux. In the Gallium experiments where main
contribution comes from 
the $pp$-neutrinos the depth of variations 
of the Ge production rate due to 
variations of the Beryllium neutrino flux is about 30\%~\cite{BFL}. 
Present statistics is  not enough to establish these variations. 
Similar situation is in  the Chlorine experiment. 
These variations can be established  by the BOREXINO \cite{BFL}.\\

Let us consider now the scenario with {\it bi-large mixing}  
(fig. \ref{bilarge}) where the solar neutrino data are explained by the
LMA
MSW conversion of 
$\nu_e$ to combination of $\nu_{\mu}$ and $\nu_{\tau}$. 
The fit of the total rates and the Night-Day asymmetry gives the following
range of the oscillation parameters: 
\begin{equation}
\Delta m^2 = (1.4 - 18) \cdot 10^{-5} {\rm eV}^2, ~~~ 
\sin^2 2\theta = 0.60 - 0.97.~~~~~~~ (3\sigma)
\end{equation}
Recent data contain some  indications in favor 
of the LMA solution~\cite{BKS99}:

\begin{figure}[htb]
\hbox to \hsize{\hfil\epsfxsize=7cm\epsfbox{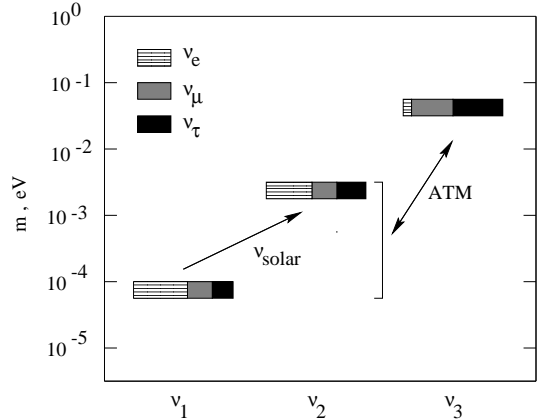}\hfil}
\caption{~~Neutrino masses and mixing pattern  of the  bi-large
mixing  scheme.
}
\label{bilarge}
\end{figure}

1). Solution predicts flat distortion of the spectrum with survival
probability determined by vacuum mixing: $P = \sin^2 \theta$. 
For $\Delta m^2 > 5 \cdot 10^{-5}~{\rm eV}^2$  the turn up of spectrum 
is expected at low energies due to effect of the adiabatic 
edge of the suppression pit. For  
$\Delta m^2 < 3 \cdot 10^{-5}~{\rm eV}^2$ small positive slope  appears
due to the Earth  regeneration effect. 
The excess of events in  the high energy
part of the spectrum, if confirmed, can be explained by the $hep$ -
neutrino  contribution~\cite{hep}.  

The Earth matter regeneration leads to  several related effects. 

2). The Night-Day asymmetry. The asymmetry increases with decrease of 
$\Delta m^2$: 
\begin{equation} 
A_{N/D} \approx  0.22~ \frac{10^{-5} {\rm eV}^2}{\Delta m^2}~, 
\end{equation}
and the dependence of $A_{N/D}$ 
 on the mixing angle is rather weak. Recent ($1\sigma$) data 
can be explained if~\cite{BKS99}  
\begin{equation}
\Delta m^2 = \left(3.5^{+2.0}_{-1.5}\right) \cdot ~10^{-5} {\rm eV}^2~. 
\end{equation}

\begin{figure}[htb]
\hbox to \hsize{\hfil\epsfxsize=10cm\epsfbox{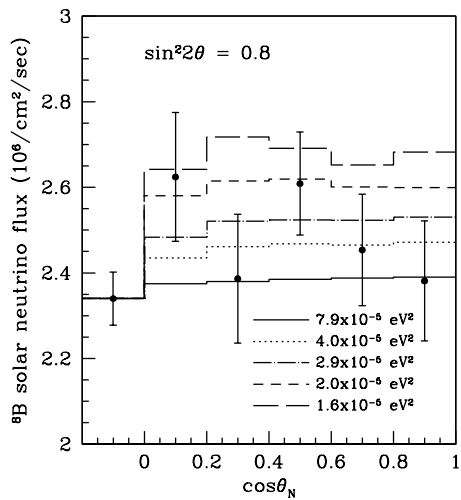}\hfil}
\caption{~~The zenith angle dependence of the total event rate 
above 6.5 MeV in the SuperKamiokande detector.  Here $\Theta_N = \pi -
\Theta$ is the nadir angle. Also shown are the SK data from
708 days. 
}
\label{zenith}
\end{figure}

3). The zenith angle distribution of events (averaged over the year) 
can lead to unique identification of the solution. 
The SK data  show that the excess of events is not concentrated in the
vertical night 
bin N5;  the excess is  observed already in the first night bin N1,  so
that the data indicate flat distribution. 
The LMA solution  predicts flat distribution for 
$\Delta m^2 > 2 \cdot 10^{-5}$ eV$^2$ (fig. \ref{zenith}). Indeed, 
for these $\Delta m^2$ the
oscillation length in matter (being of the order of the vacuum oscillation
length) is much smaller than the diameter of the Earth: 
\begin{equation}
l_m < l^{res} \equiv \frac{l_{\nu}}{\sin 2\theta}   
= \frac{4 \pi E}{\Delta m^2 \sin 2\theta} \ll d_{earth}~. 
\end{equation}
Therefore  integration over the zenith angle bin 
leads to averaging the oscillations. 
For smaller $\Delta m^2$: 
$\Delta m^2 < 1.5 \cdot 10^{-5}$ eV$^2$ and also smaller 
$\theta$  the oscillation length becomes 
larger,  
averaging is not complete and the structure appears in the 
zenith angle distribution. This however happens in the range 
of parameters  where the Night-Day
asymmetry is large (practically excluded by experiment). 

4) Seasonal asymmetry. Almost constant signal during the night 
 allows one to establish simple relation  between the 
Night-Day asymmetry (\ref{dn}) and the seasonal asymmetry (\ref{season}).  
The asymmetry appears because the nights are longer in the
winter than in the summer. 
Let $F_N$ and $F_D$ be  the constant fluxes during the night 
and during the day. Consider the integrated flux (signal) during 
 $t_0 = 24$ hours.  In winter  this averaged signal equals 
\begin{equation}
W = F_{N} t_W + F_{D} (t_0 - t_W)~;  
\end{equation}
in summer: 
\begin{equation}
S = F_{N} t_S + F_{D} (t_0 - t_S)~, 
\end{equation}
where $t_W$ ($t_S$) is the length of the night in Winter (Summer). 
Inserting  these two expressions in to (\ref{season}) and using 
the definition (\ref{dn}) 
we get after simple algebra: 
\begin{equation}
A_{W/S} = A_{N/D} \frac{t_W - t_S}{t_0}. 
\label{correl}
\end{equation}  
Thus,  the seasonal variations are proportional to the Night-Day
asymmetry. 
For the SK location the time factor in  
(\ref{correl}) is about 1/6. Thus the seasonal asymmetry due to 
regeneration is expected to be 
6 times weaker than the N/D asymmetry. For $A_{N/D} = (3 - 9)\%$ 
which corresponds to  $1\sigma$ range of 
the observed N/D effect we find $A_{W/S}  = (0.5 - 1.5)$ \%~\cite{BKS99}. 

There are several features of the seasonal asymmetry which will allow 
one to distinguish LMA regeneration case from other effects~\cite{BKS99}.

(i) No seasonal variation of the day rate is expected in contrast with 
Vacuum Oscillation case. The same is also correct for the night rate.

(ii) The regeneration asymmetry and the geometrical effect 
are in phase in the northern hemisphere and they are in opposite phase  
in the southern hemisphere which leads to cancellation.  
Thus,  the detector in southern hemisphere should
see weaker (up to factor 1/2) seasonal variations than the detector in the 
northern hemisphere.\\

\begin{figure}[htb]
\hbox to \hsize{\hfil\epsfxsize=6cm\epsfbox{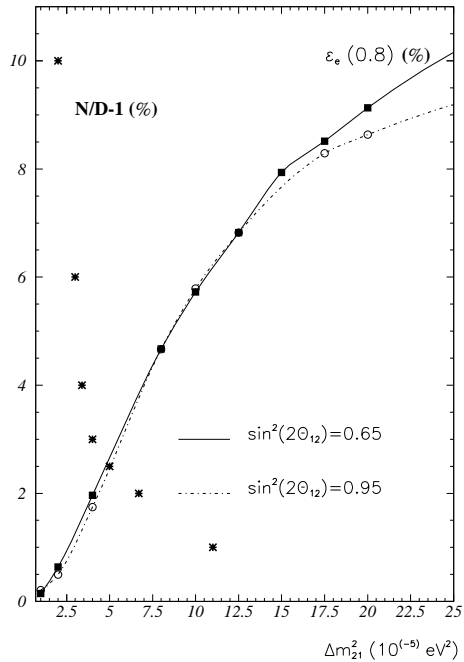}\hfil}
\caption{~~The excess of the $e$-like events as the 
function of solar $\Delta m^2$ for two different 
values of $\sin^2 2\theta_{12}$ relevant for solar neutrinos 
and  $\sin^2 2\theta_{23} = 0.8$. Also shown by crosses is the dependence 
of the Night-Day asymmetry on $\Delta m^2$. 
}
\label{excess2}
\end{figure}

Independent test of the LMA  solution follows from studies of atmospheric
neutrinos~\cite{peres}. Indeed,  the 
$\Delta m^2$ responsible for the LMA MSW solution can give an observable
effect: the $\nu_{\mu} \leftrightarrow \nu_e$-oscillations 
driven by solar $\Delta m^2$ lead to the excess of $e$-like 
atmospheric neutrino events.    
The excess can be defined as 
\begin{equation}
\epsilon_e \equiv \frac{N_e}{N_e^0} - 1~, 
\end{equation} 
where $N_e$ and $N_e^0$ are numbers of events with and without 
oscillations. 
Notice that for $\Delta m^2$ in the range of the LMA solution 
the matter suppresses the depth of oscillations (effective mixing). 
The suppression weakens with increase of $\Delta m^2$,  and 
correspondingly, 
the excess increases. In the allowed region of parameters 
it can reach $\sim 10 \%$. There is a complementarity of searches for  the
N/D  asymmetry of the solar neutrino signal and the excess of the 
$e$-like events in atmospheric neutrinos. 
According to fig. \ref{excess2} (from \cite{peres1}) with increase of
$A_{N/D}$ the excess decreases and 
vice versa. For the central value of the present N/D asymmetry the excess
is rather small: $(2 - 5) \%$. 
The excess increases with $|\cos \Theta|$,  leading to a positive
up - down 
asymmetry. The asymmetry is very weak in the low energy part of the 
sub-GeV sample ($p  < 0.4$ GeV)
but it is clearly seen in the high energy part  ($p  >  0.4$ GeV)
of the sample~\cite{peres}. The excess decreases with increase of energy
of events: 
it is practically absent in the multi-GeV sample. 
New high precision and high statistics experiments are needed to 
establish  this effect.

One more remark:  the LMA solution implies the mass of the second neutrino 
in the range $(0.4 - 1.0) \cdot 10^{-2}$ eV which is only  one order
of magnitude  smaller than the mass $m_3$ relevant for the atmospheric
neutrino anomaly: 
$$
m_2/m_3 \sim 10^{-1}. 
$$ 
All other solutions of the solar neutrino problem
require  stronger mass hierarchy. A weak hierarchy indicates 
large mixing. Indeed, the mass matrix with above ratio of the eigenstates 
can naturally lead to mixing angle $\theta_{\nu}$: 
\begin{equation}    
\tan \theta_{\nu} \sim \sqrt{\frac{m_2}{m_3}}~. 
\end{equation}
This results in  $\theta_{\nu} =  (18 - 25)^{o}$. Diagonalization of
the charge lepton matrix can give $\theta_{cl} \sim 
\sqrt{\frac{m_{\mu}}{m_{\tau}}} = (10 - 13)^{o}$, so that the total
lepton mixing 
can be $\theta_{l} = \theta_{\nu} + \theta_{cl} 
\approx (28 - 40)^{o}$ -- 
close to maximal mixing. That is,  the required large mixing
 can be obtained without special arrangements.  

\subsection{Threefold Maximal Mixing} 

In such a scheme \cite{three} moduli of all the elements of the mixing
matrix are
assumed to be equal: $|U_{\alpha i}| = 1/\sqrt{3}$. 
All three frequencies of oscillations contribute
to  all flavor channels equally.
In vacuum $\nu_{\mu} \leftrightarrow \nu_{e}$ and
$\nu_{\mu} \leftrightarrow \nu_{\tau}$ oscillations proceed 
with equal depth: $\sin^2 2\theta = 4/9$, so that the $\nu_{\mu}$ -  
disappearance is characterized by $\sin^2 2\theta = 8/9$.
Then the  CHOOZ bound implies that $\Delta m^2 < 10^{-3}$ eV$^2$.

The solar neutrino survival probability
equals $P = 4/9 P_2 + 1/9$, where $P_2$ is the two
neutrino oscillation probability with maximal depth and
smallest mass splitting $\Delta m_{12}^2$.
It is assumed that     $\Delta m_{12}^2 < 10^{-11}$ eV$^2$,
so that  1 - 2 subsystem of neutrinos is ``frozen" and $P_2 = 1$.
As the result,  the solar neutrino flux has 
energy independent suppression $P = 5/9$.

In medium with large enough density 
$V \gg \Delta m^{2}_{atm}/2E$ the channels of oscillations with $\nu_e$ 
are strongly suppressed, and the oscillation pattern is 
reduced to $\nu_{\mu} \leftrightarrow \nu_{\tau}$ oscillations 
with maximal depth. (Matter does not influence this system 
and since $\nu_{\mu}$ and  $\nu_{\tau}$ enter the scheme symmetrically 
their mixing becomes maximal after decoupling of $\nu_e$.) 
This picture is a good approximation for the multi-GeV events and 
small $\Delta m^{2}_{atm}$. However,  significant deviations
from this picture  appears for
sub-GeV sample, where $\nu_e$ channels turn out to be  suppressed weakly. 
As the consequence, one expects significant excess of the $e$-like 
events in the sub-GeV range. The analysis shows that the threefold mixing
scheme can not be excluded 
by atmospheric neutrino data and CHOOZ result at the level stronger than 
$2\sigma$ ~\cite{fogli1} (final CHOOZ result can however change this
number). 
In this scheme a description of
the solar neutrino data is rather poor, 
unless one excludes the  Homestake result from analysis. \\

\section{Maximal}

In the strict bi-maximal scheme one has $U_{e3} = 0$~\cite{bimax}. 
For large enough 
$\Delta m^{2}$ the average survival probability 
for solar neutrinos  is
about 1/2. It was marked recently~\cite{guth} that the Earth regeneration
effect is non-zero even in the case of maximal mixing which 
leads to the Night-Day effect and to dependence of the survival
probability on energy, and consequently,  to distortion of the  
recoil electron spectrum measured at the SK \cite{guth}. The point is that 
the state which arrives at the earth is pure $\nu_2$ state. 
Although this state has equal admixtures of 
the $\nu_e$ and $\nu_{\mu}$ and do not oscillate in vacuum 
it will oscillate in matter of the Earth  since the eigenstates in medium 
do not coincide with mass states 
leading to non-trivial regeneration effect. 
The  effect will be also in general  case 
of incoherent fluxes of $\nu_1$  and $\nu_2$, provided 
that their fluxes are different.  
The regeneration effect is absent only in the case of 
equal incoherent fluxes of $\nu_1$  and $\nu_2$. 

Reasonable  description of the data can be obtained if the
Homestake result is not  included in the  analysis. \\

For $\Delta m^{2}$ in the range of VO solution the data disfavor maximal
mixing which would lead to too strong distortion of the 
recoil electron spectrum. 

Clearly, the strict maximal mixing implies certain symmetry 
which is difficult to implement in view of situation with 
quark masses and mixing and charge lepton masses. \\

\section{Summary}

With high confidence level we can say  that the atmospheric 
neutrinos oscillate  and these oscillations are due to 
non-zero neutrino mass. 
High statistics  studies of the atmospheric and solar neutrinos 
will allow us to make  significant step in reconstruction 
of the neutrino mass and flavor spectrum. 
The key elements of our analysis as far as atmospheric neutrinos are
concerned,  are 

- identification of channel of neutrino oscillation; 

- clarification of the role of the electron neutrinos.  

Of course,  further checks of the oscillation 
interpretation of the 
data will be continued.\\

2. Situation with solar neutrinos is still rather uncertain. 
Main goal -- the identification of the solution is  
not yet reached. 
Different datasets  favor different solutions 
and we still should select among vacuum oscillations,  
large mixing angle  MSW, small mixing angle MSW  solution etc.. 

The most important observations which can  lead to breakthrough 
are 

- Further measurements of the N/D asymmetry and zenith angle 
distribution of events. Here it will be important  to   
include the  information  
from the low energy bins 5.5 - 6.5 MeV and then 5.0 - 5.5 MeV. 

- Measurements of spectrum distortion. Crucial  results can be obtained
from already operating SNO. 

- KamLAND~\cite{KAMLAND} experiment will allow to check LMA solution; 

- BOREXINO~\cite{BOREXINO} result can play decisive  role in
identification
of solution.\\

3. Confirmation of the LSND anomaly would be the  evidence 
of existence of sterile neutrinos (or light singlet fermion) 
which eventually can lead us  far beyond the Standard Model.


\section*{References}     

\end{document}